\documentclass[aps,twocolumn,amsmath,amssymb]{revtex4}

\usepackage{latexsym}
 
\usepackage{graphicx} 
\usepackage{color} 
\usepackage{hyperref}
\usepackage[T1]{fontenc} 
\hypersetup{
  colorlinks   = true,  
  urlcolor     = blue,  
  linkcolor    = blue,  
  citecolor    = red    
}

\begin{document} 
\bibliographystyle{apsrev} 

\title{Numerically exact approach to few-body problems far from a perturbative regime}

\author{Marcin P\l{}odzie\'n, Dariusz Wiater, Andrzej Chrostowski, Tomasz Sowi\'nski}
\affiliation{
Institute of Physics, Polish Academy of Sciences, \\ Aleja Lotnikow 32/46, PL-02668 Warsaw, Poland
}
 
\begin{abstract}

Recent developments of experimental techniques in the field of ultra-cold gases open a path to study the crossover from 'few' to 'many' on the quantum level. In this case, accurate description of inter-particle correlations is very important since it is believed that they can be utilized by quantum engineers in quantum metrology, quantum thermometry, quantum heat engines, {\it etc}. Unfortunately, a theoretical description of these correlations is very challenging since they are far beyond any variational approaches. By contrast, the exact many-body description rapidly hits numerical limitations due to an exponential increase of the many-body Hilbert space. In this work, we brush up a very effective method of constructing a many-body basis which originates in the physical argumentation. We show that, in contrast to the commonly used approach of a straightforward cut-off, it enables one to perform exact calculations with very limited numerical resources. As examples, we study quantum correlations in systems of spinless bosons and two-component mixtures of fermions confined in a one-dimensional harmonic trap being far from the perturbative regime.

\end{abstract}
 
\maketitle
 
\section{Introduction}
Amazing progress in the field of the quantum engineering proved that ultra-cold atomic systems may serve as dedicated simulators for the fundamental problems of strongly correlated quantum matter \cite{LewensteinBook}. One of the possible paths of exploration is to study ultra-cold interacting particles (bosons or fermions) confined in quasi-one-dimensional traps with they number precisely controlled \cite{Paredes2004tonks,Kinoshita2004observation,Guan2013,zurn2013Pairing,haller2009realization,Murmann2015AntiferroSpinChain,Kaufman2015Entangling,serwane2011deterministic,wenz2013fewToMany}. In this case, a general motivation is to perform a systematic and accurate analysis of the region where strong collective behavior of a few particles undergoes a specific transition to the macroscopic many-body behavior \cite{Blume2012Rev,Zinner2016Rev}. This transition seems to be crucial for further development of the quantum technologies, since it is believed that specific properties of the quantum system in this mesoscopic regime may be utilized for quantum thermometry \cite{Mann2014,Correa2017,Campbell2018},  quantum engines \cite{Kim2011,Li2012,Zhuang2014,Bracken2014,Watanabe2017,Bengtsson2018}, 
or quantum metrology \cite{Giovannetti2004,Giovannetti2011,Beau2017}. In fact, theoretical studies of correlated few-body systems are very demanding since there is a limited number of tools enabling one to perform accurate calculations. One of the most natural and widely used approaches is based on a straightforward diagonalization of the corresponding many-body Hamiltonian. In the simplest case, the Hamiltonian is represented in the many-body basis constructed from a given set of single-particle orbitals. However, when larger numbers of particles are considered, this construction becomes very ineffective and consequently, the results converge to exact ones very slowly. As shown in the seminal works \cite{Haugset1998,Deuretzbacher2007}, in the case of bosons an alternative construction of the many-body basis, based on energetic arguments, can be adopted to study properties of systems with large number of particles and very small interactions (mean-field limit) \cite{Haugset1998}, or with small number of particles and relatively strong interactions \cite{Deuretzbacher2007}. 

In our work, we widely extend this idea and we use the approach proposed in \cite{Haugset1998,Deuretzbacher2007} to predict different single- and two-particle properties in a wide range of interactions and number of particles. We systematically study the convergence of the method and compare the results with those obtained via straightforward cut-off on a single-particle basis. In addition, we go beyond bosonic systems and we systematically adopt the approach to mixtures of several fermions (repulsive and attractive). In consequence, we obtain not only ground-state energy but also a very accurate determination of correlations between fermions in systems with up to $20$ particles. Specifically, in the fermionic case, we show that the construction of the many-body basis proposed significantly reduces numerical resources needed to perform very accurate calculations. 

The paper is organized as follows. In Section~II, we discuss a general problem of the many-body basis cut-off and explain why the standard approach may lead to inaccurate results. We extend this observation in Section~III where we study properties of a few interacting bosons confined in a harmonic trap. We show that in the standard approach the amount of numerical resources is tremendous. Subsequently, in the framework of a new approach, we discuss its convergence and we predict single- and two-particle properties of the system. In Section~IV we generalize the method to the problem of a few interacting fermions and we determine specific correlations emerging for attractive interactions. Finally, in Section~V we give some additional explanations and present some numerical arguments showing that the approach for the many-body basis used is very hard to be improved. Therefore, it should be treated as the best possible implementation from the physical point of view. We conclude the paper in Section~VI.

\section{Cut-off of the basis}
Arbitrary state of interacting many-body system can be represented as a specific superposition of the many-body Fock states $\{|\mathcal{F}_k\rangle\}$. Typically, these states are constructed from the single-particle orbitals $\varphi_i(\boldsymbol{r})$ which are solutions of the corresponding single-particle Schr\"odinger equation of noninteracting particles. Depending on the quantum statistics of considered particles, the states $\{|\mathcal{F}_k\rangle\}$ encode automatically appropriate commutation relations. For example, in the case of indistinguishable and spinless fermions (bosons) given Fock state $\{|\mathcal{F}_k\rangle\}$ is constructed as a Slater determinant (permanent) of an appropriate set of $N$ orbitals. In more complicated situations (mixtures of different spices, unpolarized particles, {\it etc.}) a construction of the Fock basis is technically more complicated but it is still straightforward. This construction of Fock states has a very convenient property -- the states ${|\mathcal{F}_k\rangle}$ are automatically the eigenstates of the many-body Hamiltonian of the noninteracting system $\hat{\cal H}_0$. Although the description of the noninteracting system is very simple, the situation becomes challenging whenever mutual interactions $\hat{\cal H}_\mathtt{int}$ between particles enter the game and cannot be neglected. In principle, the eigenstates of the many-body Hamiltonian $\hat{\cal H}=\hat{\cal H}_0 + \hat{\cal H}_\mathtt{int}$ can be decomposed in the basis $\{|\mathcal{F}_k\rangle\}$ and then the problem reduces to finding the appropriate set of decomposition coefficients. 

One of the simplest and the most intuitive ways of obtaining these coefficients (at least in the case of the ground-state and several states with the lowest energy) is to perform numerically exact diagonalization of the Hamiltonian $\hat{\cal H}$ in a reduced, finite-size Hilbert space spanned by the selected Fock states. Commonly, this selection is performed from the single-particle point of view, {\it i.e.}, one selects $M$ the lowest single-particle orbitals $\varphi_i(\boldsymbol{r})$ and constructs from them all possible Fock states 
$\{|\mathcal{F}_k\rangle\}$. Consequently, in this arbitrary basis, the many-body Hamiltonian $\hat{\cal H}$ is be represented as a matrix which can be diagonalized
This approach is based on the assumption that whenever cut-off point $M$ is increased, obtained eigenstates become closer to the true eigenstates of the Hamiltonian $\hat{\cal H}$. In practice, the convergence of the method is quite slow and it is effective only for the lowest eigenstates of the system and not for very strong interactions.  

Although, the method of cutting-off the single-particle basis is very intuitive and straightforward, it {\it is essentially not} systematic from the physical point of view. As noted in \cite{Haugset1998}, in the many-body language it takes into account states with relatively high energy, neglecting in the same time other states with energy evidently smaller. In consequence, it unnecessarily induces uncertainties of the final results and it inevitably leads to a waste of huge amount of numerical resources. This observation can be utilized to find the much more accurate construction of an appropriate basis in the many-body Hilbert space. In consequence, with the same amount of numerical resources, one can find eigenstates of an interacting many-body system and their measurable properties with much larger accuracy. Alternatively, keeping the same accuracy one can study the systems with a much larger number of particles and/or larger interaction strengths.  

\section{Ultra-cold bosons in a harmonic trap}
The observation outlined above is very general and it   can be applied almost to any many-body system. However, to make our presentation as clear as possible, let us first focus on the one of the simplest nontrivial cases of a few bosons of mass $m$, confined in one-dimensional harmonic trap of the frequency $\Omega$, and interacting via point-like interactions \cite{Deuretzbacher2007,Zollner2007,koscik2012quantum,GarciaMarch2014,Harshman2016I,Harshman2016II}. In this case the model Hamiltonian of the system can be written in the second quantization form as following:
\begin{align} \label{Hamiltonian}
\hat{\cal H} &= \int\!\!\mathrm{d}x\,\hat\Psi^\dagger(x)\left(-\frac{\hbar^2}{2m}\frac{\mathrm{d}^2}{\mathrm{d}x^2}+\frac{m\Omega^2}{2}x^2\right)\hat{\Psi}(x) \nonumber \\
&+\int\!\!\mathrm{d}x\!\int\!\!\mathrm{d}x'\,\hat\Psi^\dagger(x)\hat\Psi^\dagger(x'){\cal V}(x-x')\hat{\Psi}(x')\hat{\Psi}(x)
\end{align}
where the field operator $\hat\Psi(x)$ annihilates boson at point $x$ and satisfies standard commutation relations $[\Psi(x),\Psi^\dagger(x')]=\delta(x-x')$ and $[\Psi(x),\Psi(x')]=0$. Since we consider system in the ultra-cold regime it is quite good approximation to model inter-particle forces with zero-range potential $V(r)=g\delta(r)$. Note, that in one-dimension (in contrast to higher dimensions) any regularization of the $\delta$-like potential is not needed since it is well defined hermitian and self-adjoint operator.  In this case, the single-particle orbitals $\varphi_i(x)$ are simply given by standard harmonic oscillator wave functions
\begin{equation}
\varphi_i(x) = {\cal N}_i\,\mathrm{exp}\left(-\frac{x^2}{2\lambda^2}\right)\mathbf{H}_i\left(\frac{x}{\lambda}\right),
\end{equation}
where $\lambda=\sqrt{\hbar/m\Omega}$ is a natural oscillator length, ${\cal N}_i$ is a normalization factor, and the functions $\mathbf{H}_i(\xi)$ are the Hermite polynomials. In this case the single-particle energies are simply given by $\epsilon_i=\hbar\Omega(i+1/2)$. Typically, one expands the field operator $\hat\Psi(x)$ in this basis 
\begin{equation} \label{Decomp}
\hat\Psi(x) = \sum_{i=0}^\infty \varphi_i(x) \hat{a}_i
\end{equation}
and introduces a bosonic operator $\hat{a}_i$ annihilating particle described by a single-particle state $\varphi_i(x)$. Corresponding Fock basis $\{|{\cal F}_k\rangle\}$ of $N$ bosons is spanned by the following vectors
\begin{equation} \label{StandardFock}
|{\cal F}_k\rangle \equiv |n_1, n_2, \ldots \rangle \sim (\hat{a}_1^\dagger)^{n_1}\cdot(\hat{a}_2^\dagger)^{n_2}\cdots|\mathtt{vac}\rangle
\end{equation}
where index $k$ enumerates consecutive distributions of $N$ particles in single-particle states and $n_i$ is the number of bosons occupying the state described by $\varphi_i(x)$. Obviously, occupation numbers satisfy a constrain $\sum_i n_i = N$. The ground-state of the noninteracting system is represented by the Fock State $|{\cal F}_0\rangle=|N,0,\ldots\rangle$ with the energy $E_0=N/2$. Energy of other Fock states can be calculated straightforwardly as $E_k=\sum_i \epsilon_i n_i$.

\begin{figure} 
\includegraphics[scale=1.6]{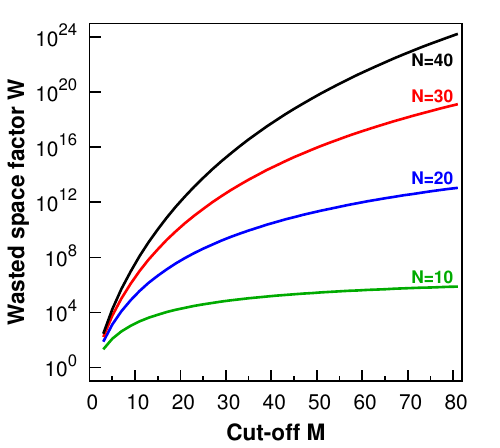}
\caption{Wasted space factor $\mathbf{W}$ defined according to the eq.~\eqref{WFactor} as a function of the single-particle cut-off $M$ for a different number of bosons. Note a logarithmic scale of the vertical axis. \label{Fig1}}
\end{figure}

As explained previously, due to the numerical limitations, in the standard approach, one cuts-off a size of the single-particle basis on some large but finite number $M$, {\it i.e.}, the summation in \eqref{Decomp} runs from $0$ to $M-1$. Then, the Hilbert space is spanned by the finite number of Fock states and, in the considered case of indistinguishable bosons, its dimension is given by
\begin{equation}
  \mathbf{D}(M,N) = \frac{(M+N-1)!}{(M-1)!\, N!}.
\end{equation}
Note that in the Hilbert space with the cut-off $M$ the states with minimal and maximal single-particle energy are represented by the Fock states $|N,0,\ldots,0\rangle$ and $|0,\ldots,0,N\rangle$, respectively. The energy of the later state is equal $E_{max}=N(M-1/2)$. In fact, the latter state is the only state having this energy in the cropped Hilbert space. However, this is no longer true when the whole physical Hilbert space of infinite dimension is considered. For example, the Fock state with one particle promoted to the state $M+1$, one particle relegated to the state $M-1$, and with $N-2$ particles remaining in the state $M$ has evidently the same energy. However, due to unphysical cut-off introduced to perform numerical analysis, this state and plenty of other states with the same energy are not taken into account at all. The same story can be said about states with other energies represented only partially in the cropped Hilbert space. In fact, for a given cut-off $M$ only the energies not larger than $E_{opt}=(N-1)/2+(M-1/2)$ (corresponding to the state $|N-1,0,\ldots,0,1\rangle$) are appropriately represented in the cropped Hilbert space. Let us denote the number of these states as $\mathbf{d}(M,N)$. All states with larger energies are taken into account inconsequently and their number is tremendously large when compared to $\mathbf{d}$. Especially, when a large number of particles and large cut-offs $M$ are considered. 
To show how huge is an amount of numerical resources wasted due to the inconsistent choice of Fock states in Fig.~\ref{Fig1} we show how the wasted space factor
\begin{equation} \label{WFactor}
\mathbf{W}(M,N) = \frac{\mathbf{D}(M,N)-\mathbf{d}(M,N)}{\mathbf{d}(M,N)}
\end{equation}
depends on cut-off $M$ and the number of particles $N$ (note a logarithmic scale on vertical axis). It is quite obvious that along with increasing cut-off a number of inappropriately selected states in the Hilbert space groves tremendously. 
From the physical point of view, there is no reason to favor these states over these states with the same or lower energy which was neglected due to a technical procedure of cutting the single-particle basis. This fact has direct implications for practical calculations of physical quantities with different numerical methods. 

\begin{figure}
\includegraphics[scale=0.9]{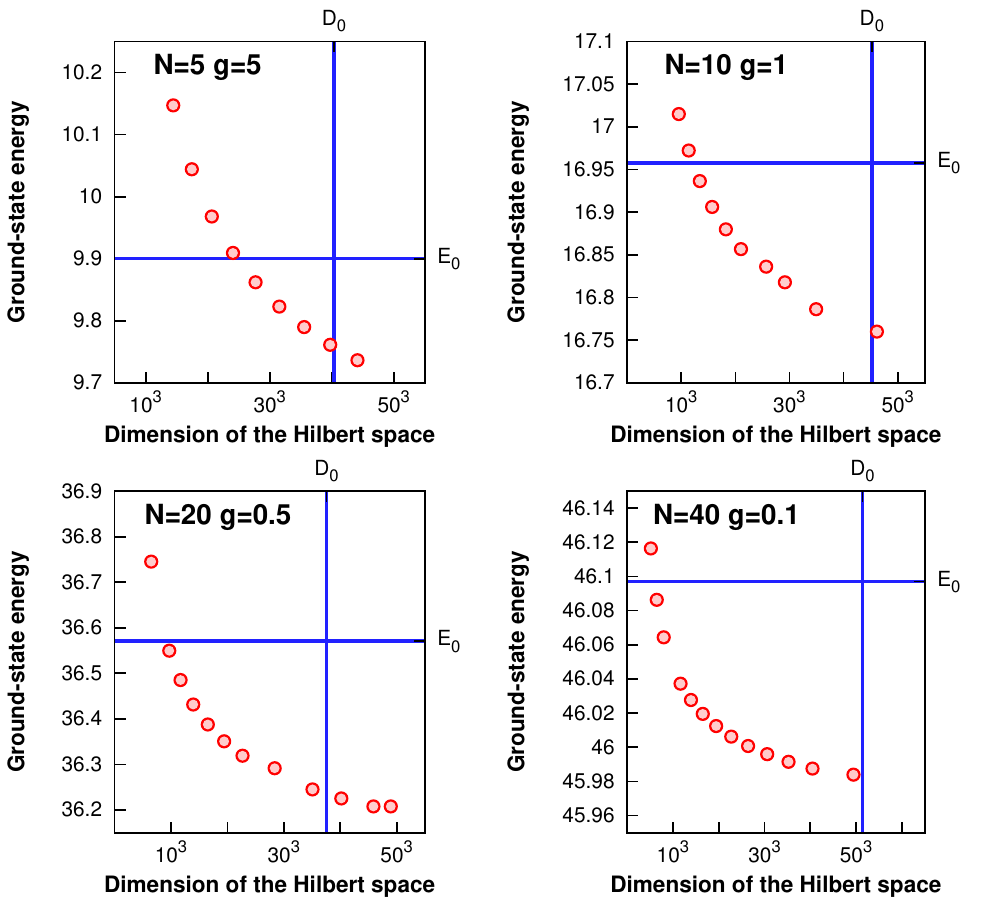}
\caption{Energy of the ground-state of $N$ bosons as a function of a the Hilbert space size obtained with the exact diagonalization of the many-body Hamiltonian \eqref{Hamiltonian} for a different number of particles and different interaction strengths. Blue horizontal line indicates the ground-state energy $E_0$ obtained with standard single-particle cut-off when the dimension of the Hilbert space is equal to $D_0$ (vertical blue line). Red dots represent energies obtained with the improved method of the many-body energy cut-off (see the main text for details). It is clear that an accuracy of the improved approach is significantly better for the same numerical resources. Note the nonlinear scale on the horizontal axis to increase visibility. Ground-state energy is measured in natural harmonic oscillator units $\hbar\Omega$.\label{Fig2}}
\end{figure}

Instead of cutting the Fock basis with respect to the single-particle orbitals one can construct the many-body basis by selecting Fock states with consecutive energies. In the case of a harmonic oscillator, it can be done straightforwardly since the problem of finding all Fock states of a given energy is equivalent to the mathematical problem of finding all possible partitions of a corresponding integer \cite{Sandor2006Partitions}. For other confinements the situation is much more demanding, but still, the effective algorithm for generating the basis exists. To show that this approach in fact significantly increases an accuracy of the results (for given numerical resources available) let us concentrate on the simplest quantity of the many-body system, {\it i.e.}, its ground-state energy. In Fig.~\ref{Fig2} we plot this quantity obtained with exact diagonalization of the many-body Hamiltonian \eqref{Hamiltonian} for a different number of particles and different interactions. Blue vertical and horizontal lines indicate the size of the Hilbert space $D_0$ and corresponding ground-state energy $E_0$ obtained with a straightforward construction of the many-body basis states \eqref{StandardFock} based on single-particle cut-off. In contrast, red dots corresponds to similar calculations performed in the basis formed by states having consecutive energies. It is clearly seen that with the latter method the ground-state energy is obtained with much higher accuracy when the size of the basis become equal to $D_0$. Moreover, an accuracy of the standard approach (the horizontal line indicating energy $E_0$) is achieved for significantly smaller basis constructed with respect to energy levels. For example, for $N=10$ particles, the energy $E_0$ is achieved with $D_0=92\,378$ states in standard procedure based on single-particle cut-off. The same energy is obtained with only $d=2\,430$ states when the energy cut-off approach is adopted. 

At this point it is worth to mention an additional advantage of the improved method. In the standard approach, when we increase the cut-off by one single-particle state we observe an inflation of the corresponding size of the many-body Hilbert. For example, for $N=20$ and $M=6$ one finds $D_0=53\,130$ while for $M=7$ $D_0=230\,230$. It simply means that in practice it is not possible to perform more heavy calculations and shift vertical blue lines to the next cut-off position since the corresponding Hilbert space is extremely large. In fact, the effect is unmanageable for a larger number of particles and therefore the method is simply useless. In the method based on the energy cut-off, an increase of the energy on which we cut the basis is less influential to the size of the space. Therefore, it is much easier to perform systematic calculations and test a convergence of the results. It is clearly visible on Fig.~\ref{Fig2}.  

\begin{figure}
\includegraphics[scale=0.9]{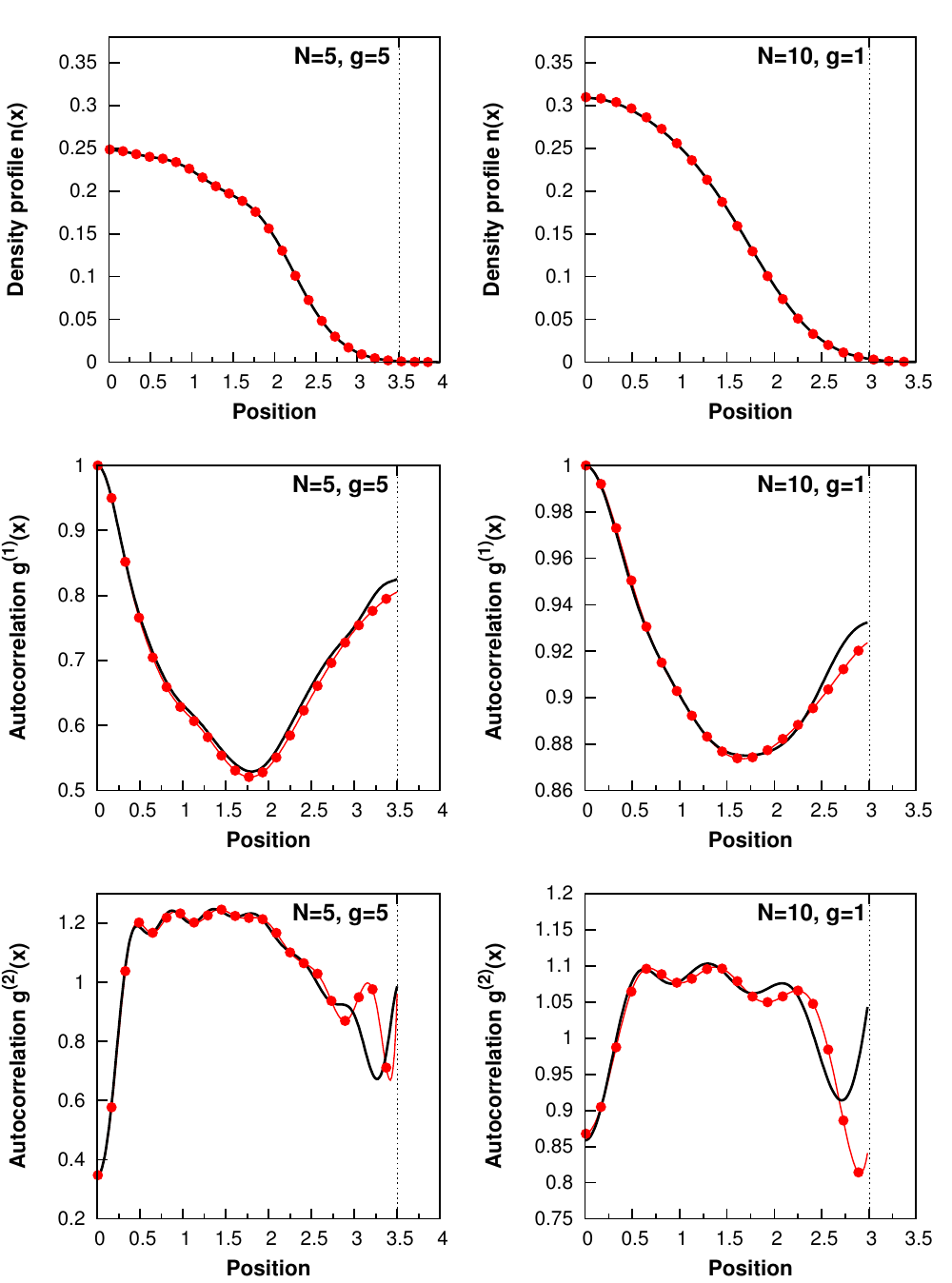}
\caption{Properties of the many-body ground-state of a few bosons in the regime of strong repulsions obtained with two complementary methods: standard single-particle cut-off (thin black line) and improved cut-off based on energy arguments (thick red line). Vertical dotted lines mark positions where exponential vanishing of the density occurs. Note that although there is a nice agreement between methods for density profiles, discrepancies are clearly visible when correlations in the system are considered. See the main text for details. All positions and the density profile are measured in natural units of harmonic oscillator $\lambda$ and $\lambda^{-1}$, respectively. Autocorrelation functions $g^{(1)}$ and $g^{(2)}$ are dimensionless.\label{Fig3}}
\end{figure}

Obviously, deviations between the two methods become significant when the system is far from the perturbative regime and higher single-particle orbitals start to contribute in the ground state of the system.  Particularly it is visible for larger number of particles when other experimentally accessible quantities than the ground-state energy are considered. The simplest quantities of this kind are related to different single-particle measurements and they are encoded in the reduced single-particle density matrix of the form
\begin{equation}
\rho^{(1)}(x;x') = \frac{1}{N}\langle \mathtt{G}_0|\hat{\Psi}^\dagger(x)\hat{\Psi}(x')|\mathtt{G}_0\rangle.
\end{equation}
Typically, one is interested in its diagonal form $n(x)=\rho^{(1)}(x;x)$ encoding the single-particle density profile and the off-diagonal autocorrelation function $g^{(1)}(x)$ defined as
\begin{equation}
g^{(1)}(x) = \frac{\rho^{(1)}(-x;x)}{\rho(x)}
\end{equation}
which measures the single-particle long-range order in the ground-state. To determine non-classical correlations between particles also some two-particle correlations are considered. The simplest quantity of this kind, typically used when ultra-cold bosons are studied, is the two-body autocorrelation function \begin{equation}
g^{(2)}(x)=\frac{\rho^{(2)}(x,-x;x,-x)}{\left[\rho(x)\right]^2}
\end{equation}
where $\rho^{(2)}$ is the reduced two-particle density matrix
\begin{multline}
\rho^{(2)}(x_1,x_2;x_1',x_2') = \\ \frac{1}{N(N\!-\!1)}\langle \mathtt{G}_0|\hat{\Psi}^\dagger(x_1)\hat{\Psi}^\dagger(x_2)\hat{\Psi}(x_2')\hat{\Psi}(x_1')|\mathtt{G}_0\rangle.
\end{multline}

In Fig.~\ref{Fig3} we plot these different quantities for two different numbers of particles in a strong interaction regime. Two lines correspond to the results obtained with standard single-particle cut-off method (thin black line) and the energy cut-off approach (dotted red line), respectively. For clearness, we cut down both autocorrelation plots to regions of non-vanishing densities (marked with dotted vertical line). Red dotted lines are obtained for relatively small numerical resources ($d=12\,519$ for $N=5$ and $d=1\,136$ for $N=10$). We treat these results as being very close to the exact values since they are almost not sensitive when the many-body basis is increased. In contrast, the results obtained with straightforward cut of single-particle basis are obtained for much larger resources ($D=65\,780$ for $N=5$ and $D=43\,758$ for $N=10$). Moreover, in the case of autocorrelation functions $g^{(1)}$ and $g^{(2)}$, they cannot be treated as numerically converged since they change significantly when the cut-off $M$ is increased. By performing precise calculations with an enlarged amount of numerical resources we checked that resulting curves from the standard single-particle cut-off method (black lines) slowly converge to those obtained with cut-off with respect to the many-body energy (dotted red lines). All these suggest that the standard cut-off method on a single-particle basis should be used with a particular attention when correlations between particles in a strong interaction regime are considered.

\section{Two-flavored mixture of fermions}
To show that the method of cutting the Fock basis with respect to the many-body energy may significantly increase accuracy of numerical predictions of different many-body systems let us now concentrate on a two-component mixture of a few fermions confined in a harmonic trap. Different properties of these systems were recently extensively studied theoretically \cite{Sowinski2013,Gharashi2013UpperBranchCorrelations,bugnion2013ferromagnetic,Yan2014Temperature,Lindgren2014Fermionization,Sowinski2015Pairing,DAmico2015Pairing,YangPRA2015,Sowinski2018FreeSpin}. For simplicity, we assume only contact interactions between particles however considerations for other interactions can be performed analogously. In this case the Hamiltonian of the system has the following form
\begin{align} \label{HamFerm}
\hat{\cal H} &= \sum_\sigma\int\!\!\mathrm{d}x\,\hat\Psi_\sigma^\dagger(x)\left(-\frac{\hbar^2}{2m}\frac{\mathrm{d}^2}{\mathrm{d}x^2}+\frac{m\Omega^2}{2}x^2\right)\hat{\Psi}_\sigma(x) \nonumber \\
&+\int\!\!\mathrm{d}x\,\hat\Psi_\downarrow^\dagger(x)\hat\Psi_\uparrow^\dagger(x)\hat{\Psi}_\uparrow(x)\hat{\Psi}_\downarrow(x),
\end{align}
where $\hat{\Psi}_\sigma(x)$ is a fermionic field operator corresponding to the component $\sigma\in\{\uparrow,\downarrow\}$ and it obeys anti-commutation relations, $\{\hat\Psi_\sigma(x),\hat\Psi_{\sigma'}^\dagger(x')\}=\delta_{\sigma\sigma'}\delta(x-x')$ and $\{\hat\Psi_\sigma(x),\hat\Psi_{\sigma'}(x')\}=0$. Due to these relations the wave function of the many-body system has to be antisymmetrized under exchange of any two particles. In consequence the zero-range part of any mutual interaction vanishes for fermions belonging to the same component. In this case a decomposition of the field operator in a basis of single-particle orbitals has a form
\begin{equation}
\hat{\Psi}_\sigma(x)=\sum_i \hat{b}_{\sigma i}\,\varphi_i(x)
\end{equation}
where $\hat{b}_{\sigma i}$ is a fermionic operator anihilating particle with spin $\sigma$ in a state $\varphi_i(x)$. Since the Hamiltonian \eqref{HamFerm} commutes with operators counting numbers of particles in a given spin $\hat{N}_\sigma = \sum_i \hat{b}^\dagger_{\sigma i}\hat{b}_{\sigma i}$, therefore a whole analysis can be performed in subspaces of given  distribution of particle among components. For a system of $N=N_\uparrow+N_\downarrow$ particles, the corresponding many-body Hilbert space is spanned by Fock states constructed as
\begin{multline}
|{\cal F}_k\rangle \equiv |n_1, n_2, \ldots;m_1, m_2, \ldots\rangle \\ \sim (\hat{b}_{\uparrow 1}^\dagger)^{n_1}\cdot(\hat{b}_{\uparrow 2}^\dagger)^{n_2}\cdots(\hat{b}_{\downarrow 1}^\dagger)^{n_1}\cdot(\hat{b}_{\downarrow 2}^\dagger)^{n_2}\cdots|\mathtt{vac}\rangle,
\end{multline}
where $\sum_i n_i = N_\uparrow$ and $\sum_i m_i = N_\downarrow$ and $n_i,m_i\in\{0,1\}$. If the Hilbert space is cut on some single-particle basis $M$ then the many-body space is spanned by a finite number of Fock states and its dimension is given by
\begin{equation}
\mathbf{D}(M,N_\uparrow,N_\downarrow) = \frac{M!}{(M-N_\uparrow)!\,N_\uparrow!}\cdot\frac{M!}{(M-N_\downarrow)!\,N_\downarrow!}.
\end{equation}
The state with the lowest energy (corresponding to the non-interacting many-body ground-state) has the energy equal to the Fermi energy $E_{F}=(N_\uparrow^2+N_\downarrow^2)/2$, while the highest excited state in the cropped Hilbert space with cut-off $M$ has the energy $E_{max}=M(N_\uparrow+N_\downarrow)-E_{F}$. Similarly to the bosonic case, cutting the Hilbert space with respect to the single-particle orbitals directly leads to inconsistent consideration of states with higher energies. It can be shown that in the case of fermionic mixtures the optimal energy $E_{opt}$ is given by
\begin{equation} 
E_{opt} = M+E_F-\mathrm{max}(N_\uparrow,N_\downarrow),
\end{equation}
{\it i.e.}, all energies larger than $E_{opt}$ are inappropriately represented in the cropped Hilbert space \cite{Generalization}. Consequently, huge amount of numerical resources is unnecessarily wasted. 
\begin{figure}
\includegraphics[width=\linewidth]{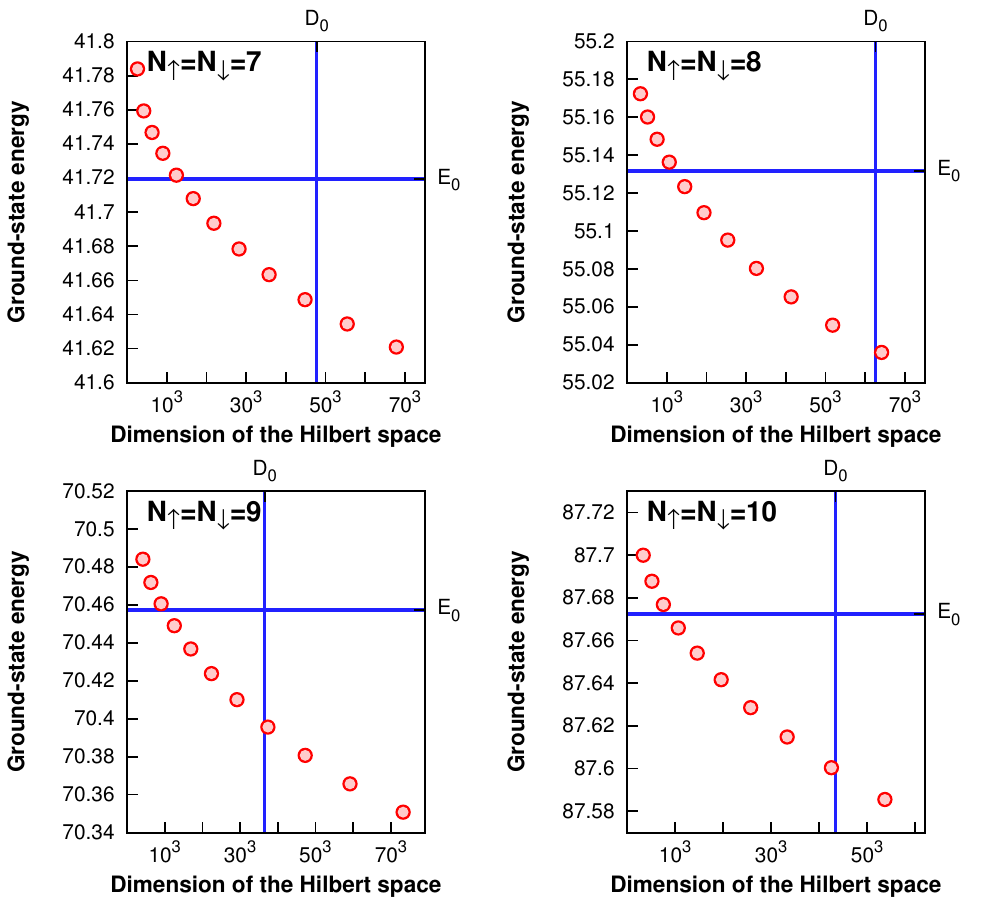}
\caption{Energy of the ground-state of attractively interacting fermions ($g=-1$) as a function of the Hilbert space determined by available numerical resources to perform exact diagonalization. Blue horizontal line indicates the ground-state energy $E_0$ obtained with standard single-particle cut-off when the dimension of the Hilbert space is equal $D_0$ (vertical blue line). Red dots represent energies obtained with improved method of tha many-body energy cut-off (see the main text for details). Similarly to the bosonic case, an accuracy of the improved approach is significantly better for the same numerical resources. Note the nonlinear scale on vertical axis to increase visibility. Ground-state energy is measured in natural harmonic oscillator units $\hbar\Omega$. \label{Fig4}}
\end{figure}

\begin{figure}
\includegraphics[width=\linewidth]{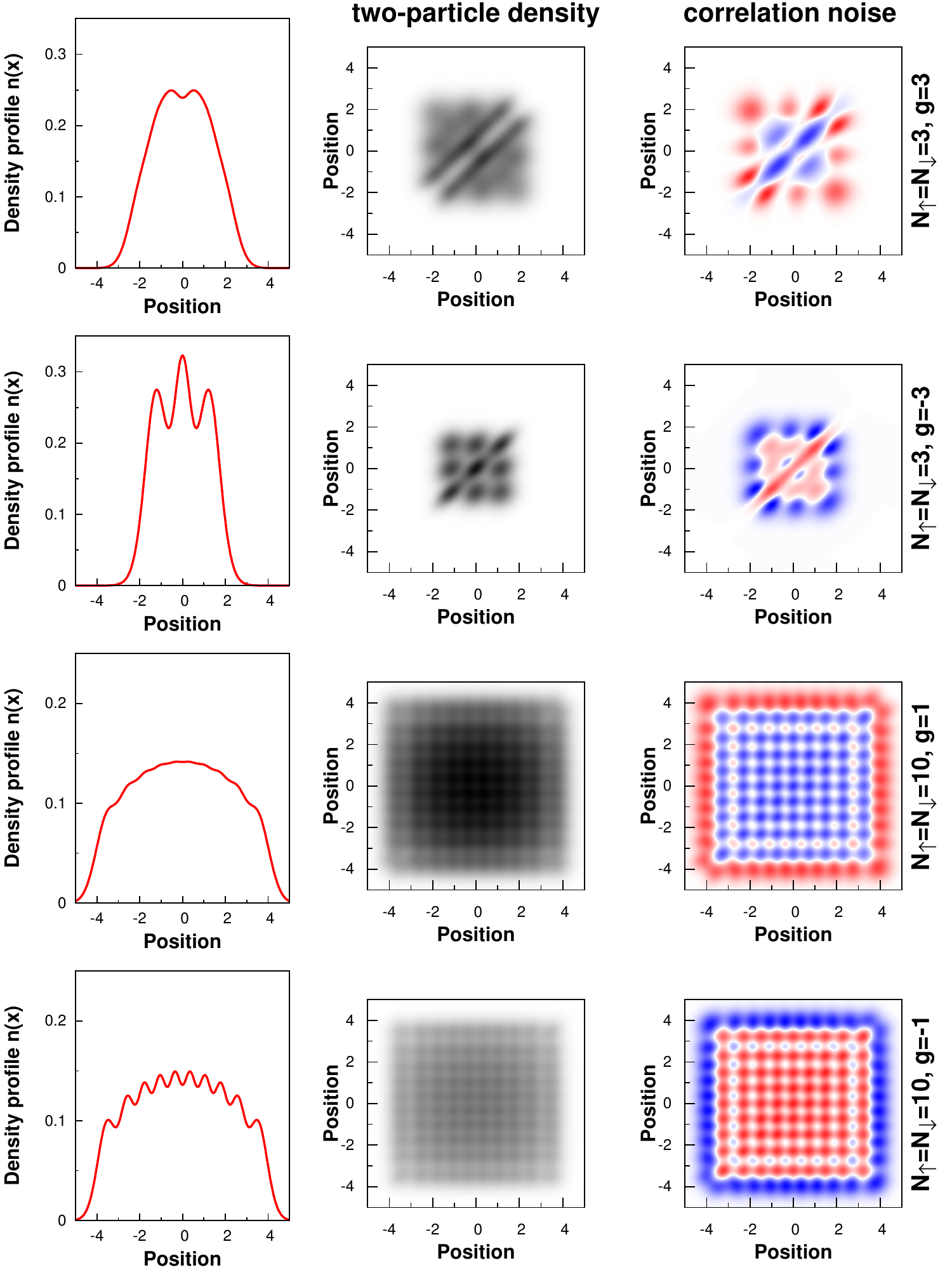}
\caption{ Different properties of the many-body ground-state of a mixture of a few interacting fermions. Successive rows correspond to different numbers of particles and different interactions $N_\uparrow=N_\downarrow=3$, $g=\pm 3$ and $N_\uparrow=N_\downarrow=10$, $g=\pm 1$, respectively. In columns we show single-particle density profile \eqref{FermDensity}, two-particle density \eqref{FermPairDensity}, and correlation noise density \eqref{FermNoise} (see the main text for details). All results are well converged and they are obtained via exact diagonalization in the Fock basis cut with respect to the many-body energy. Positions and single-particle density profiles are measured in natural units of harmonic oscillator $\lambda$ and $\lambda^{-1}$, respectively. Two-particle density and correlation noise are measured in $\lambda^{-2}$.\label{Fig5}}
\end{figure}

Having all this information in mind, one can perform calculations in a full analogy to the bosonic case by 
filling the many-body Fock basis with states from consecutive energy shells. By performing an exact diagonalization of the many-body Hamiltonian in this basis accuracies of the resulting eigenstates and corresponding eigenenergies are significantly improved. For example, in Fig.~\ref{Fig4} we plot convergence of the ground-state energy of attractively interacting ($g=-1$) fermionic mixture for a different number of particles. As in the bosonic case, having numerical resources fixed (in our case an ability to diagonalize matrices with sizes less than $3\times10^5$) we converge much closer to the ground-state than in the standard approach based on single-particle cut-off. Having this very accurate approximation of the many-body ground state one can find its different properties which are far beyond capabilities of the standard cut-off method. The simplest is the single-particle density profile of \textcolor{blue}{a} given component
\begin{equation} \label{FermDensity}
n_\sigma (x) = \frac{1}{N_\sigma}\langle \mathtt{G}_0|\hat{\Psi}_\sigma^\dagger(x)\hat{\Psi}_\sigma(x)|\mathtt{G}_0\rangle.
\end{equation}
In the first column in Fig.~\ref{Fig5} we show this quantity calculated for repulsive, as well as for attractive interactions far from the perturbative regime, {\it i.e.}, for relatively small number of particles $N_\uparrow=N_\downarrow=3$ and strong interactions $|g|=3$, or for large particle number $N_\uparrow=N_\downarrow=10$ and interactions $|g|=1$. In all these cases the results are well converged, {\it i.e.}, they are insensitive to further enlargement of the Fock basis. For attractive forces a characteristic modulation of the density profile is visible. 

The simplest quantity which describes relative relations between different components is the two-particle density distribution (middle column in Fig.~\ref{Fig5}) defined as
\begin{multline} \label{FermPairDensity}
\rho (x;y) =  \frac{1}{N_\downarrow N_\uparrow}\langle \mathtt{G}_0|\hat{\Psi}_\downarrow^\dagger(x)\hat{\Psi}_\uparrow^\dagger(y)\hat{\Psi}_\uparrow(y)\hat{\Psi}_\downarrow(x)|\mathtt{G}_0\rangle.
\end{multline}
As it is seen, the joint probability of finding two opposite fermions is position dependent and some characteristic density patterns are visible. Note also that in the case of attractive forces and strong interactions (second row in Fig.~\ref{Fig5}) strong correlations in positions are visible and probability of finding both particles exactly in the same place is dominant. In this regime also some strong anti-correlation in momenta of opposite-spin fermions is present (not shown here) which is understood as a first predictor of Cooper pairing phenomena \cite{zurn2013Pairing,Sowinski2015Pairing,DAmico2015Pairing}.

Two-particle density distribution has one fundamental limitation when correlations between quantum particles are discussed. Namely, it does not discriminate between correlations induced by mutual interactions and coincidental meeting of two uncorrelated particles. To have better discrimination of these two elements one introduces a concept of the spatial correlation noise \cite{Brandt2017}. It is defined as
\begin{equation} \label{FermNoise}
{\cal G} (x;y) = \rho(x;y) - n_\downarrow(x)n_\uparrow(y).
\end{equation}
Simply, this two-particle distribution vanishes at given position whenever the two-particle density profile can be viewed as a simple product of single-particle distributions of corresponding components. As it is seen in the third column of Fig.~\ref{Fig5}, the correlation noise distribution emphasizes quantum correlations between particles very clearly. Different regions of positive and negative correlation noise are visible. Since the correlation noise measures deviations from the uncorrelated two-particle density distribution, therefore there is some numerical challenge to get results well-converged, especially for systems being far from the perturbative regime. Due to an appropriate numerical approach based on energy arguments for selecting elements of Fock basis we are able to predict almost exactly the correlation noise up to 20 particles and unfold correlations with a very compelling spatial distribution (bottom rows in Fig.~\ref{Fig5}).

\section{Final remarks}
\begin{figure}
\includegraphics[width=\linewidth]{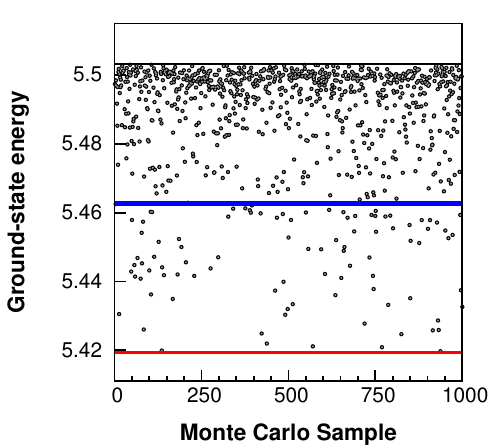}
\caption{ 
Ground-state energy for $N = 5$ interacting ($g=1$) bosons in a harmonic trap. Blue (middle) horizontal line corresponds to the energy obtained with standard single-particle cut-off method $M=10$ (Size of the Hilbert space $D_0=2002$). If the Fock basis od the same size is build with many-body states of consecutive energies then the ground-state energy is significantly improved (red horizontal line). Dots present ground-state energies obtained when the Fock basis is built with randomly chosen many-body states. See the main text for details.
\label{Fig6}}
\end{figure}

Actually, the method based on the energy cut-off of the basis has a very simple and straightforward physical interpretation based on energetic arguments. It is quite obvious that corrections to a selected eigenstate of the many-body Hamiltonian originating in couplings to states added when numerical resources are expanded, besides quantum-mechanical amplitude calculated as an appropriate expectation value of the Hamiltonian, is instantaneously suppressed by the energy gap to these states. It means that the most important contribution comes from the states with the lowest energy. Consequently, if it is possible, one should extend the basis by states having the lowest possible energies. One can perform simple numerical argument showing that obtaining a better approach to the problem is very hard if possible at all. The example of this argumentation for $N=5$ bosons and interaction $g=1$ is illustrated in Fig.~\ref{Fig6}. The blue horizontal line represents the ground-state energy obtained for single-particle cut-off $M=10$ (size of the Hilbert space $D_0=2002$), whereas black horizontal is obtained when only $d$ basis states (those with energies not larger than $E_{opt}$) are taken into account. Note, that only these states are taken consistently with respect to the energy for assumed cut-off $M$. Instead of filling remaining resources with states of the fixed cut-off one can adapt our approach and fill them with states with consecutive energies up to the Hilbert space size $D_0$. The resulting ground-state energy obtained is represented by the red horizontal line. To show that this approach is the most efficient we also fill the remaining space with states choose completely randomly from the set of states with energies larger than $E_{opt}$ and arbitrary cut-off. Corresponding energies are represented with grey dots (values on horizontal axis correspond to consecutive random samples). As it is seen, energies obtained in this way are always larger than the energy obtained with the energy-shell approach (red line). Moreover, from this point of view, the energy obtained via single-particle cut-off is unexceptional since it has some average value which can be simply improved by a blind and random selection of states. It means that, at least in the case of a few quantum particles, the method based on cutting the basis with respect to energies is the most efficient approaches to perform numerical calculations. 

\section{Conclusions}
In this work we theoretically study correlations in few-body systems of ultra-cold bosons and mixtures of fermions in the framework of exact diagonalization approach. We show that standard method of cutting the many-body basis is highly ineffective when systems with quite a large number of particles in the non-perturbative regime are considered. To get well-converged results we adopt the method based on the energy of many-body states which enable us to perform accurate calculations with limited numerical resources. We believe that this approach opens up a next possible pathway for exact studies of collective properties of few-body problems which are extensively studied in nowadays experiments.

\section*{Acknowledgements}
The authors are very grateful to Przemys{\l}aw Ko\'scik and Mariusz Gajda for their valuable remarks and comments. This work was supported by the (Polish) National Science Center Grant No. 2016/22/E/ST2/00555.

\bibliography{_bibtex}

\begin{thebibliography}{43}
\expandafter\ifx\csname natexlab\endcsname\relax\def\natexlab#1{#1}\fi
\expandafter\ifx\csname bibnamefont\endcsname\relax
  \def\bibnamefont#1{#1}\fi
\expandafter\ifx\csname bibfnamefont\endcsname\relax
  \def\bibfnamefont#1{#1}\fi
\expandafter\ifx\csname citenamefont\endcsname\relax
  \def\citenamefont#1{#1}\fi
\expandafter\ifx\csname url\endcsname\relax
  \def\url#1{\texttt{#1}}\fi
\expandafter\ifx\csname urlprefix\endcsname\relax\def\urlprefix{URL }\fi
\providecommand{\bibinfo}[2]{#2}
\providecommand{\eprint}[2][]{\url{#2}}

\bibitem[{\citenamefont{Lewenstein et~al.}(2012)\citenamefont{Lewenstein,
  Sanpera, and Ahufinger}}]{LewensteinBook}
\bibinfo{author}{\bibfnamefont{M.}~\bibnamefont{Lewenstein}},
  \bibinfo{author}{\bibfnamefont{A.}~\bibnamefont{Sanpera}}, \bibnamefont{and}
  \bibinfo{author}{\bibfnamefont{V.}~\bibnamefont{Ahufinger}},
  \emph{\bibinfo{title}{{Ultracold Atoms in Optical Lattices: Simulating
  quantum many-body systems}}} (\bibinfo{publisher}{Oxford University Press},
  \bibinfo{address}{Oxford}, \bibinfo{year}{2012}).

\bibitem[{\citenamefont{Paredes et~al.}(2004)\citenamefont{Paredes, Widera,
  Murg, Mandel, F{\"o}lling, Cirac, Shlyapnikov, H{\"a}nsch, and
  Bloch}}]{Paredes2004tonks}
\bibinfo{author}{\bibfnamefont{B.}~\bibnamefont{Paredes}},
  \bibinfo{author}{\bibfnamefont{A.}~\bibnamefont{Widera}},
  \bibinfo{author}{\bibfnamefont{V.}~\bibnamefont{Murg}},
  \bibinfo{author}{\bibfnamefont{O.}~\bibnamefont{Mandel}},
  \bibinfo{author}{\bibfnamefont{S.}~\bibnamefont{F{\"o}lling}},
  \bibinfo{author}{\bibfnamefont{I.}~\bibnamefont{Cirac}},
  \bibinfo{author}{\bibfnamefont{G.~V.} \bibnamefont{Shlyapnikov}},
  \bibinfo{author}{\bibfnamefont{T.~W.} \bibnamefont{H{\"a}nsch}},
  \bibnamefont{and} \bibinfo{author}{\bibfnamefont{I.}~\bibnamefont{Bloch}},
  \bibinfo{journal}{Nature} \textbf{\bibinfo{volume}{429}},
  \bibinfo{pages}{277} (\bibinfo{year}{2004}).

\bibitem[{\citenamefont{Kinoshita et~al.}(2004)\citenamefont{Kinoshita, Wenger,
  and Weiss}}]{Kinoshita2004observation}
\bibinfo{author}{\bibfnamefont{T.}~\bibnamefont{Kinoshita}},
  \bibinfo{author}{\bibfnamefont{T.}~\bibnamefont{Wenger}}, \bibnamefont{and}
  \bibinfo{author}{\bibfnamefont{D.~S.} \bibnamefont{Weiss}},
  \bibinfo{journal}{Science} \textbf{\bibinfo{volume}{305}},
  \bibinfo{pages}{1125} (\bibinfo{year}{2004}).

\bibitem[{\citenamefont{Guan et~al.}(2013)\citenamefont{Guan, Batchelor, and
  Lee}}]{Guan2013}
\bibinfo{author}{\bibfnamefont{X.~W.} \bibnamefont{Guan}},
  \bibinfo{author}{\bibfnamefont{M.~T.} \bibnamefont{Batchelor}},
  \bibnamefont{and} \bibinfo{author}{\bibfnamefont{C.}~\bibnamefont{Lee}},
  \bibinfo{journal}{Rev. Mod. Phys.} \textbf{\bibinfo{volume}{85}},
  \bibinfo{pages}{1633} (\bibinfo{year}{2013}).

\bibitem[{\citenamefont{Z\"urn et~al.}(2013)\citenamefont{Z\"urn, Wenz,
  Murmann, Bergschneider, Lompe, and Jochim}}]{zurn2013Pairing}
\bibinfo{author}{\bibfnamefont{G.}~\bibnamefont{Z\"urn}},
  \bibinfo{author}{\bibfnamefont{A.~N.} \bibnamefont{Wenz}},
  \bibinfo{author}{\bibfnamefont{S.}~\bibnamefont{Murmann}},
  \bibinfo{author}{\bibfnamefont{A.}~\bibnamefont{Bergschneider}},
  \bibinfo{author}{\bibfnamefont{T.}~\bibnamefont{Lompe}}, \bibnamefont{and}
  \bibinfo{author}{\bibfnamefont{S.}~\bibnamefont{Jochim}},
  \bibinfo{journal}{Phys. Rev. Lett.} \textbf{\bibinfo{volume}{111}},
  \bibinfo{pages}{175302} (\bibinfo{year}{2013}).

\bibitem[{\citenamefont{Haller et~al.}(2009)\citenamefont{Haller, Gustavsson,
  Mark, Danzl, Hart, Pupillo, and N{\"a}gerl}}]{haller2009realization}
\bibinfo{author}{\bibfnamefont{E.}~\bibnamefont{Haller}},
  \bibinfo{author}{\bibfnamefont{M.}~\bibnamefont{Gustavsson}},
  \bibinfo{author}{\bibfnamefont{M.~J.} \bibnamefont{Mark}},
  \bibinfo{author}{\bibfnamefont{J.~G.} \bibnamefont{Danzl}},
  \bibinfo{author}{\bibfnamefont{R.}~\bibnamefont{Hart}},
  \bibinfo{author}{\bibfnamefont{G.}~\bibnamefont{Pupillo}}, \bibnamefont{and}
  \bibinfo{author}{\bibfnamefont{H.~C.} \bibnamefont{N{\"a}gerl}},
  \bibinfo{journal}{Science} \textbf{\bibinfo{volume}{325}},
  \bibinfo{pages}{1224} (\bibinfo{year}{2009}).

\bibitem[{\citenamefont{Murmann et~al.}(2015)\citenamefont{Murmann,
  Deuretzbacher, Z\"{u}rn, Bjerlin, Reimann, Santos, Lompe, and
  Jochim}}]{Murmann2015AntiferroSpinChain}
\bibinfo{author}{\bibfnamefont{S.}~\bibnamefont{Murmann}},
  \bibinfo{author}{\bibfnamefont{F.}~\bibnamefont{Deuretzbacher}},
  \bibinfo{author}{\bibfnamefont{G.}~\bibnamefont{Z\"{u}rn}},
  \bibinfo{author}{\bibfnamefont{J.}~\bibnamefont{Bjerlin}},
  \bibinfo{author}{\bibfnamefont{S.~M.} \bibnamefont{Reimann}},
  \bibinfo{author}{\bibfnamefont{L.}~\bibnamefont{Santos}},
  \bibinfo{author}{\bibfnamefont{T.}~\bibnamefont{Lompe}}, \bibnamefont{and}
  \bibinfo{author}{\bibfnamefont{S.}~\bibnamefont{Jochim}},
  \bibinfo{journal}{Phys. Rev. Lett.} \textbf{\bibinfo{volume}{115}},
  \bibinfo{pages}{215301} (\bibinfo{year}{2015}).

\bibitem[{\citenamefont{Kaufman et~al.}(2015)\citenamefont{Kaufman, Lester,
  Foss-Feig, Wall, Rey, and Regal}}]{Kaufman2015Entangling}
\bibinfo{author}{\bibfnamefont{A.~M.} \bibnamefont{Kaufman}},
  \bibinfo{author}{\bibfnamefont{B.~J.} \bibnamefont{Lester}},
  \bibinfo{author}{\bibfnamefont{M.}~\bibnamefont{Foss-Feig}},
  \bibinfo{author}{\bibfnamefont{M.~L.} \bibnamefont{Wall}},
  \bibinfo{author}{\bibfnamefont{A.~M.} \bibnamefont{Rey}}, \bibnamefont{and}
  \bibinfo{author}{\bibfnamefont{C.~A.} \bibnamefont{Regal}},
  \bibinfo{journal}{Nature} \textbf{\bibinfo{volume}{527}},
  \bibinfo{pages}{208} (\bibinfo{year}{2015}).

\bibitem[{\citenamefont{Serwane et~al.}(2011)\citenamefont{Serwane, Z{\"u}rn,
  Lompe, Ottenstein, Wenz, and Jochim}}]{serwane2011deterministic}
\bibinfo{author}{\bibfnamefont{F.}~\bibnamefont{Serwane}},
  \bibinfo{author}{\bibfnamefont{G.}~\bibnamefont{Z{\"u}rn}},
  \bibinfo{author}{\bibfnamefont{T.}~\bibnamefont{Lompe}},
  \bibinfo{author}{\bibfnamefont{T.}~\bibnamefont{Ottenstein}},
  \bibinfo{author}{\bibfnamefont{A.}~\bibnamefont{Wenz}}, \bibnamefont{and}
  \bibinfo{author}{\bibfnamefont{S.}~\bibnamefont{Jochim}},
  \bibinfo{journal}{Science} \textbf{\bibinfo{volume}{332}},
  \bibinfo{pages}{336} (\bibinfo{year}{2011}).

\bibitem[{\citenamefont{Wenz et~al.}(2013)\citenamefont{Wenz, Z{\"u}rn,
  Murmann, Brouzos, Lompe, and Jochim}}]{wenz2013fewToMany}
\bibinfo{author}{\bibfnamefont{A.}~\bibnamefont{Wenz}},
  \bibinfo{author}{\bibfnamefont{G.}~\bibnamefont{Z{\"u}rn}},
  \bibinfo{author}{\bibfnamefont{S.}~\bibnamefont{Murmann}},
  \bibinfo{author}{\bibfnamefont{I.}~\bibnamefont{Brouzos}},
  \bibinfo{author}{\bibfnamefont{T.}~\bibnamefont{Lompe}}, \bibnamefont{and}
  \bibinfo{author}{\bibfnamefont{S.}~\bibnamefont{Jochim}},
  \bibinfo{journal}{Science} \textbf{\bibinfo{volume}{342}},
  \bibinfo{pages}{457} (\bibinfo{year}{2013}).

\bibitem[{\citenamefont{Blume}(2012)}]{Blume2012Rev}
\bibinfo{author}{\bibfnamefont{D.}~\bibnamefont{Blume}}, \bibinfo{journal}{Rep.
  Prog. Phys.} \textbf{\bibinfo{volume}{75}}, \bibinfo{pages}{046401}
  (\bibinfo{year}{2012}).

\bibitem[{\citenamefont{Zinner}(2016)}]{Zinner2016Rev}
\bibinfo{author}{\bibfnamefont{N.~T.} \bibnamefont{Zinner}},
  \bibinfo{journal}{EPJ Web of Conferences} \textbf{\bibinfo{volume}{113}},
  \bibinfo{pages}{01002} (\bibinfo{year}{2016}).

\bibitem[{\citenamefont{Mann and Mart{\'\i}n-Mart{\'\i}nez}(2014)}]{Mann2014}
\bibinfo{author}{\bibfnamefont{R.~B.} \bibnamefont{Mann}} \bibnamefont{and}
  \bibinfo{author}{\bibfnamefont{E.}~\bibnamefont{Mart{\'\i}n-Mart{\'\i}nez}},
  \bibinfo{journal}{Found. Phys.} \textbf{\bibinfo{volume}{44}},
  \bibinfo{pages}{492} (\bibinfo{year}{2014}).

\bibitem[{\citenamefont{Correa et~al.}(2017)\citenamefont{Correa,
  Perarnau-Llobet, Hovhannisyan, Hern\'andez-Santana, Mehboudi, and
  Sanpera}}]{Correa2017}
\bibinfo{author}{\bibfnamefont{L.~A.} \bibnamefont{Correa}},
  \bibinfo{author}{\bibfnamefont{M.}~\bibnamefont{Perarnau-Llobet}},
  \bibinfo{author}{\bibfnamefont{K.~V.} \bibnamefont{Hovhannisyan}},
  \bibinfo{author}{\bibfnamefont{S.}~\bibnamefont{Hern\'andez-Santana}},
  \bibinfo{author}{\bibfnamefont{M.}~\bibnamefont{Mehboudi}}, \bibnamefont{and}
  \bibinfo{author}{\bibfnamefont{A.}~\bibnamefont{Sanpera}},
  \bibinfo{journal}{Phys. Rev. A} \textbf{\bibinfo{volume}{96}},
  \bibinfo{pages}{062103} (\bibinfo{year}{2017}).

\bibitem[{\citenamefont{Campbell et~al.}(2018)\citenamefont{Campbell, Genoni,
  and Deffner}}]{Campbell2018}
\bibinfo{author}{\bibfnamefont{S.}~\bibnamefont{Campbell}},
  \bibinfo{author}{\bibfnamefont{M.~G.} \bibnamefont{Genoni}},
  \bibnamefont{and} \bibinfo{author}{\bibfnamefont{S.}~\bibnamefont{Deffner}},
  \bibinfo{journal}{Quantum Sci. Technol.} \textbf{\bibinfo{volume}{3}},
  \bibinfo{pages}{025002} (\bibinfo{year}{2018}).

\bibitem[{\citenamefont{Kim et~al.}(2011)\citenamefont{Kim, Sagawa,
  De~Liberato, and Ueda}}]{Kim2011}
\bibinfo{author}{\bibfnamefont{S.~W.} \bibnamefont{Kim}},
  \bibinfo{author}{\bibfnamefont{T.}~\bibnamefont{Sagawa}},
  \bibinfo{author}{\bibfnamefont{S.}~\bibnamefont{De~Liberato}},
  \bibnamefont{and} \bibinfo{author}{\bibfnamefont{M.}~\bibnamefont{Ueda}},
  \bibinfo{journal}{Phys. Rev. Lett.} \textbf{\bibinfo{volume}{106}},
  \bibinfo{pages}{070401} (\bibinfo{year}{2011}).

\bibitem[{\citenamefont{Li et~al.}(2012)\citenamefont{Li, Zou, Li, Shao, and
  Wu}}]{Li2012}
\bibinfo{author}{\bibfnamefont{H.}~\bibnamefont{Li}},
  \bibinfo{author}{\bibfnamefont{J.}~\bibnamefont{Zou}},
  \bibinfo{author}{\bibfnamefont{J.-G.} \bibnamefont{Li}},
  \bibinfo{author}{\bibfnamefont{B.}~\bibnamefont{Shao}}, \bibnamefont{and}
  \bibinfo{author}{\bibfnamefont{L.-A.} \bibnamefont{Wu}},
  \bibinfo{journal}{Annals of Physics} \textbf{\bibinfo{volume}{327}},
  \bibinfo{pages}{2955 } (\bibinfo{year}{2012}).

\bibitem[{\citenamefont{Zhuang and Liang}(2014)}]{Zhuang2014}
\bibinfo{author}{\bibfnamefont{Z.}~\bibnamefont{Zhuang}} \bibnamefont{and}
  \bibinfo{author}{\bibfnamefont{S.-D.} \bibnamefont{Liang}},
  \bibinfo{journal}{Phys. Rev. E} \textbf{\bibinfo{volume}{90}},
  \bibinfo{pages}{052117} (\bibinfo{year}{2014}).

\bibitem[{\citenamefont{Bracken}(2014)}]{Bracken2014}
\bibinfo{author}{\bibfnamefont{P.}~\bibnamefont{Bracken}},
  \bibinfo{journal}{Cent. Eur. J. Phys.} \textbf{\bibinfo{volume}{12}},
  \bibinfo{pages}{1} (\bibinfo{year}{2014}).

\bibitem[{\citenamefont{Watanabe et~al.}(2017)\citenamefont{Watanabe,
  Venkatesh, Talkner, and del Campo}}]{Watanabe2017}
\bibinfo{author}{\bibfnamefont{G.}~\bibnamefont{Watanabe}},
  \bibinfo{author}{\bibfnamefont{B.~P.} \bibnamefont{Venkatesh}},
  \bibinfo{author}{\bibfnamefont{P.}~\bibnamefont{Talkner}}, \bibnamefont{and}
  \bibinfo{author}{\bibfnamefont{A.}~\bibnamefont{del Campo}},
  \bibinfo{journal}{Phys. Rev. Lett.} \textbf{\bibinfo{volume}{118}},
  \bibinfo{pages}{050601} (\bibinfo{year}{2017}).

\bibitem[{\citenamefont{Bengtsson et~al.}(2018)\citenamefont{Bengtsson,
  Tengstrand, Wacker, Samuelsson, Ueda, Linke, and Reimann}}]{Bengtsson2018}
\bibinfo{author}{\bibfnamefont{J.}~\bibnamefont{Bengtsson}},
  \bibinfo{author}{\bibfnamefont{M.~N.} \bibnamefont{Tengstrand}},
  \bibinfo{author}{\bibfnamefont{A.}~\bibnamefont{Wacker}},
  \bibinfo{author}{\bibfnamefont{P.}~\bibnamefont{Samuelsson}},
  \bibinfo{author}{\bibfnamefont{M.}~\bibnamefont{Ueda}},
  \bibinfo{author}{\bibfnamefont{H.}~\bibnamefont{Linke}}, \bibnamefont{and}
  \bibinfo{author}{\bibfnamefont{S.~M.} \bibnamefont{Reimann}},
  \bibinfo{journal}{Phys. Rev. Lett.} \textbf{\bibinfo{volume}{120}},
  \bibinfo{pages}{100601} (\bibinfo{year}{2018}).

\bibitem[{\citenamefont{Giovannetti et~al.}(2004)\citenamefont{Giovannetti,
  Lloyd, and Maccone}}]{Giovannetti2004}
\bibinfo{author}{\bibfnamefont{V.}~\bibnamefont{Giovannetti}},
  \bibinfo{author}{\bibfnamefont{S.}~\bibnamefont{Lloyd}}, \bibnamefont{and}
  \bibinfo{author}{\bibfnamefont{L.}~\bibnamefont{Maccone}},
  \bibinfo{journal}{Science} \textbf{\bibinfo{volume}{306}},
  \bibinfo{pages}{1330} (\bibinfo{year}{2004}).

\bibitem[{\citenamefont{Giovannetti et~al.}(2011)\citenamefont{Giovannetti,
  Lloyd, and Maccone}}]{Giovannetti2011}
\bibinfo{author}{\bibfnamefont{V.}~\bibnamefont{Giovannetti}},
  \bibinfo{author}{\bibfnamefont{S.}~\bibnamefont{Lloyd}}, \bibnamefont{and}
  \bibinfo{author}{\bibfnamefont{L.}~\bibnamefont{Maccone}},
  \bibinfo{journal}{Nature Photonics} \textbf{\bibinfo{volume}{5}},
  \bibinfo{pages}{222} (\bibinfo{year}{2011}).

\bibitem[{\citenamefont{Beau and del Campo}(2017)}]{Beau2017}
\bibinfo{author}{\bibfnamefont{M.}~\bibnamefont{Beau}} \bibnamefont{and}
  \bibinfo{author}{\bibfnamefont{A.}~\bibnamefont{del Campo}},
  \bibinfo{journal}{Phys. Rev. Lett.} \textbf{\bibinfo{volume}{119}},
  \bibinfo{pages}{010403} (\bibinfo{year}{2017}).

\bibitem[{\citenamefont{Haugset and Haugerud}(1998)}]{Haugset1998}
\bibinfo{author}{\bibfnamefont{T.}~\bibnamefont{Haugset}} \bibnamefont{and}
  \bibinfo{author}{\bibfnamefont{H.}~\bibnamefont{Haugerud}},
  \bibinfo{journal}{Phys. Rev. A} \textbf{\bibinfo{volume}{57}},
  \bibinfo{pages}{3809} (\bibinfo{year}{1998}).

\bibitem[{\citenamefont{Deuretzbacher et~al.}(2007)\citenamefont{Deuretzbacher,
  Bongs, Sengstock, and Pfannkuche}}]{Deuretzbacher2007}
\bibinfo{author}{\bibfnamefont{F.}~\bibnamefont{Deuretzbacher}},
  \bibinfo{author}{\bibfnamefont{K.}~\bibnamefont{Bongs}},
  \bibinfo{author}{\bibfnamefont{K.}~\bibnamefont{Sengstock}},
  \bibnamefont{and}
  \bibinfo{author}{\bibfnamefont{D.}~\bibnamefont{Pfannkuche}},
  \bibinfo{journal}{Phys. Rev. A} \textbf{\bibinfo{volume}{75}},
  \bibinfo{pages}{013614} (\bibinfo{year}{2007}).

\bibitem[{\citenamefont{Z\"ollner et~al.}(2007)\citenamefont{Z\"ollner, Meyer,
  and Schmelcher}}]{Zollner2007}
\bibinfo{author}{\bibfnamefont{S.}~\bibnamefont{Z\"ollner}},
  \bibinfo{author}{\bibfnamefont{H.-D.} \bibnamefont{Meyer}}, \bibnamefont{and}
  \bibinfo{author}{\bibfnamefont{P.}~\bibnamefont{Schmelcher}},
  \bibinfo{journal}{Phys. Rev. A} \textbf{\bibinfo{volume}{75}},
  \bibinfo{pages}{043608} (\bibinfo{year}{2007}).

\bibitem[{\citenamefont{Ko{\'s}cik}(2012)}]{koscik2012quantum}
\bibinfo{author}{\bibfnamefont{P.}~\bibnamefont{Ko{\'s}cik}},
  \bibinfo{journal}{Few Body Syst.} \textbf{\bibinfo{volume}{52}},
  \bibinfo{pages}{49} (\bibinfo{year}{2012}).

\bibitem[{\citenamefont{Garc\'{\i}a-March
  et~al.}(2014)\citenamefont{Garc\'{\i}a-March, Juli\'{a}-D\'{\i}az,
  Astrakharchik, Boronat, and Polls}}]{GarciaMarch2014}
\bibinfo{author}{\bibfnamefont{M.~A.} \bibnamefont{Garc\'{\i}a-March}},
  \bibinfo{author}{\bibfnamefont{B.}~\bibnamefont{Juli\'{a}-D\'{\i}az}},
  \bibinfo{author}{\bibfnamefont{G.~E.} \bibnamefont{Astrakharchik}},
  \bibinfo{author}{\bibfnamefont{J.}~\bibnamefont{Boronat}}, \bibnamefont{and}
  \bibinfo{author}{\bibfnamefont{A.}~\bibnamefont{Polls}},
  \bibinfo{journal}{Phys. Rev. A} \textbf{\bibinfo{volume}{90}},
  \bibinfo{pages}{063605} (\bibinfo{year}{2014}).

\bibitem[{\citenamefont{Harshman}(2016{\natexlab{a}})}]{Harshman2016I}
\bibinfo{author}{\bibfnamefont{N.~L.} \bibnamefont{Harshman}},
  \bibinfo{journal}{Few-Body Syst.} \textbf{\bibinfo{volume}{57}},
  \bibinfo{pages}{11} (\bibinfo{year}{2016}{\natexlab{a}}).

\bibitem[{\citenamefont{Harshman}(2016{\natexlab{b}})}]{Harshman2016II}
\bibinfo{author}{\bibfnamefont{N.~L.} \bibnamefont{Harshman}},
  \bibinfo{journal}{Few-Body Syst.} \textbf{\bibinfo{volume}{57}},
  \bibinfo{pages}{45} (\bibinfo{year}{2016}{\natexlab{b}}).

\bibitem[{San(2006)}]{Sandor2006Partitions}
\emph{\bibinfo{title}{Handbook of Number Theory I. Partitions}}
  (\bibinfo{publisher}{Springer Netherlands}, \bibinfo{address}{Dordrecht},
  \bibinfo{year}{2006}), pp. \bibinfo{pages}{491--521}, ISBN
  \bibinfo{isbn}{978-1-4020-3658-3}.

\bibitem[{\citenamefont{Sowi\'{n}ski et~al.}(2013)\citenamefont{Sowi\'{n}ski,
  Grass, Dutta, and Lewenstein}}]{Sowinski2013}
\bibinfo{author}{\bibfnamefont{T.}~\bibnamefont{Sowi\'{n}ski}},
  \bibinfo{author}{\bibfnamefont{T.}~\bibnamefont{Grass}},
  \bibinfo{author}{\bibfnamefont{O.}~\bibnamefont{Dutta}}, \bibnamefont{and}
  \bibinfo{author}{\bibfnamefont{M.}~\bibnamefont{Lewenstein}},
  \bibinfo{journal}{Phys. Rev. A} \textbf{\bibinfo{volume}{88}},
  \bibinfo{pages}{033607} (\bibinfo{year}{2013}).

\bibitem[{\citenamefont{Gharashi and
  Blume}(2013)}]{Gharashi2013UpperBranchCorrelations}
\bibinfo{author}{\bibfnamefont{S.~E.} \bibnamefont{Gharashi}} \bibnamefont{and}
  \bibinfo{author}{\bibfnamefont{D.}~\bibnamefont{Blume}},
  \bibinfo{journal}{Phys. Rev. Lett.} \textbf{\bibinfo{volume}{111}},
  \bibinfo{pages}{045302} (\bibinfo{year}{2013}).

\bibitem[{\citenamefont{Bugnion and Conduit}(2013)}]{bugnion2013ferromagnetic}
\bibinfo{author}{\bibfnamefont{P.}~\bibnamefont{Bugnion}} \bibnamefont{and}
  \bibinfo{author}{\bibfnamefont{G.}~\bibnamefont{Conduit}},
  \bibinfo{journal}{Physical Review A} \textbf{\bibinfo{volume}{87}},
  \bibinfo{pages}{060502} (\bibinfo{year}{2013}).

\bibitem[{\citenamefont{Yan and Blume}(2014)}]{Yan2014Temperature}
\bibinfo{author}{\bibfnamefont{Y.}~\bibnamefont{Yan}} \bibnamefont{and}
  \bibinfo{author}{\bibfnamefont{D.}~\bibnamefont{Blume}},
  \bibinfo{journal}{Phys. Rev. A} \textbf{\bibinfo{volume}{90}},
  \bibinfo{pages}{013620} (\bibinfo{year}{2014}).

\bibitem[{\citenamefont{Lindgren et~al.}(2014)\citenamefont{Lindgren, Rotureau,
  Forss{\'e}n, Volosniev, and Zinner}}]{Lindgren2014Fermionization}
\bibinfo{author}{\bibfnamefont{E.}~\bibnamefont{Lindgren}},
  \bibinfo{author}{\bibfnamefont{J.}~\bibnamefont{Rotureau}},
  \bibinfo{author}{\bibfnamefont{C.}~\bibnamefont{Forss{\'e}n}},
  \bibinfo{author}{\bibfnamefont{A.}~\bibnamefont{Volosniev}},
  \bibnamefont{and} \bibinfo{author}{\bibfnamefont{N.~T.}
  \bibnamefont{Zinner}}, \bibinfo{journal}{New J. Phys.}
  \textbf{\bibinfo{volume}{16}}, \bibinfo{pages}{063003}
  (\bibinfo{year}{2014}).

\bibitem[{\citenamefont{Sowi{\'n}ski et~al.}(2015)\citenamefont{Sowi{\'n}ski,
  Gajda, and Rz\c{a}{\.z}ewski}}]{Sowinski2015Pairing}
\bibinfo{author}{\bibfnamefont{T.}~\bibnamefont{Sowi{\'n}ski}},
  \bibinfo{author}{\bibfnamefont{M.}~\bibnamefont{Gajda}}, \bibnamefont{and}
  \bibinfo{author}{\bibfnamefont{K.}~\bibnamefont{Rz\c{a}{\.z}ewski}},
  \bibinfo{journal}{Europhys. Lett.} \textbf{\bibinfo{volume}{109}},
  \bibinfo{pages}{26005} (\bibinfo{year}{2015}).

\bibitem[{\citenamefont{D'Amico and Rontani}(2015)}]{DAmico2015Pairing}
\bibinfo{author}{\bibfnamefont{P.}~\bibnamefont{D'Amico}} \bibnamefont{and}
  \bibinfo{author}{\bibfnamefont{M.}~\bibnamefont{Rontani}},
  \bibinfo{journal}{Phys. Rev. A} \textbf{\bibinfo{volume}{91}},
  \bibinfo{pages}{043610} (\bibinfo{year}{2015}).

\bibitem[{\citenamefont{Yang et~al.}(2015)\citenamefont{Yang, Guan, and
  Pu}}]{YangPRA2015}
\bibinfo{author}{\bibfnamefont{L.}~\bibnamefont{Yang}},
  \bibinfo{author}{\bibfnamefont{L.}~\bibnamefont{Guan}}, \bibnamefont{and}
  \bibinfo{author}{\bibfnamefont{H.}~\bibnamefont{Pu}}, \bibinfo{journal}{Phys.
  Rev. A} \textbf{\bibinfo{volume}{91}}, \bibinfo{pages}{043634}
  (\bibinfo{year}{2015}).

\bibitem[{\citenamefont{Sowi\'nski}(2018)}]{Sowinski2018FreeSpin}
\bibinfo{author}{\bibfnamefont{T.}~\bibnamefont{Sowi\'nski}},
  \bibinfo{journal}{Condens. Matter} \textbf{\bibinfo{volume}{3}},
  \bibinfo{pages}{7} (\bibinfo{year}{2018}).

\bibitem[{Gen()}]{Generalization}
\bibinfo{note}{Expression for an optimal energy can be easily generalized to
  larger number of components. For $K$-component mixture with
  $N_1,N_2,\ldots,N_K$ fermions it has a form $E_{opt} =
  M+E_F-\mathrm{max}(N_1,N_2,\ldots,N_K)$, where Fermi energy is given by
  $E_F=\sum_k N_k^2/2$.}

\bibitem[{\citenamefont{Brandt et~al.}(2017)\citenamefont{Brandt, Yannouleas,
  and Landman}}]{Brandt2017}
\bibinfo{author}{\bibfnamefont{B.~B.} \bibnamefont{Brandt}},
  \bibinfo{author}{\bibfnamefont{C.}~\bibnamefont{Yannouleas}},
  \bibnamefont{and} \bibinfo{author}{\bibfnamefont{U.}~\bibnamefont{Landman}},
  \bibinfo{journal}{Phys. Rev. A} \textbf{\bibinfo{volume}{96}},
  \bibinfo{pages}{053632} (\bibinfo{year}{2017}).

\end{thebibliography}
\end{document}